\documentclass[useAMS,usegraphicx,usenatbib]{mn2e}
\usepackage{times}
\usepackage{url}

\title[Fe-H nanoparticles in space]{Missing Fe:  hydrogenated iron nanoparticles}
\author[G. Bilalbegovi\' c et al]
{G. Bilalbegovi\' c$^{1}$,
A. Maksimovi\' c$^{2}$,
V. Moha\v cek-Gro\v sev$^{2}$
\\
$^{1}$Department of Physics, Faculty of Science, University of Zagreb, Bijeni\v cka 32, 10000 Zagreb, Croatia\\
$^{2}$Center of Excellence for Advanced Materials and Sensing Devices, Rudjer Bo\v skovi\' c Institute, Bijeni\v  cka 54, 10000 Zagreb, Croatia 
}

\begin{document}

\date{\today}

\pubyear{2016} \volume{000}

\maketitle 
\label{firstpage}

\begin{abstract}
Although it was found that the FeH lines exist in the spectra of some stars, none of the spectral  features in the ISM have been assigned to this molecule.  We suggest that iron atoms interact with hydrogen and produce Fe-H nanoparticles which sometimes contain many H atoms.  We calculate infrared spectra of hydrogenated iron nanoparticles using density functional theory methods and find broad, overlapping bands. Desorption of H$_2$ could induce spinning of these small Fe-H dust grains. Some of hydrogenated iron nanoparticles posses magnetic and electric moments and should interact with electromagnetic fields in the ISM. Fe$_n$H$_m$  nanoparticles could contribute to the polarization of the ISM and the anomalous microwave emission. We discuss the conditions required to form FeH and Fe$_n$H$_m$ in the ISM.
\end{abstract}


\section{Introduction}
\label{intro}

Iron is one of the most abundant chemical elements on Earth and in the Galaxy.  It is proposed that more than 65$\%$ of Fe is injected as gas into the ISM \citep{Dwek2016}.
However, it is known that iron  is strongly depleted \citep{Savage1979,Jensen2007,Delgado2009,Jenkins2009,Jones2013,Dwek2016}. Therefore, it was suggested that Fe atoms hide in dust grains. 
Although the number of known molecules in the ISM approaches two hundreds \citep{Tielens2013}, only two chemical species containing iron have been detected: FeO and FeCN \citep{Endres2016,Walmsley2002,Furuya2003,Zack2011}. Therefore, the search for additional iron species in the ISM and cosmic dust is necessary.

The most abundant element in the universe is hydrogen. It is important to consider chemical compounds consisting of iron and hydrogen. The FeH molecule has been discussed in the astrophysical literature. 
It was proposed that FeH forms in the Sun and some other stars \citep{Carroll1976}. 
The spectrum of several stars in the ultraviolet, visible, and near-infrared region was compared with laboratory measurements of FeH
and good agreement for some lines was obtained.  The near-infrared  Wing-Ford band at 0.99  $\mu$m  was connected with the FeH spectral features 
in S stars, M-type giants, 
M and L dwarfs \citep{Nordh1977,Wing1977,Clegg1978,Jones1996,Kirkpatrick1999,McLean2000,Buenzli2015}. The extensive literature about FeH is available on the ExoMol web page: \url{http://www.exomol.com/bibliography/FeH} \citep{Tennyson2012}.

Studies in materials science have shown that the Fe atom and small iron clusters could bind many hydrogen atoms \citep{Whetten1985,Richtsmeier1985,Parks1985,Knickelbein1998,Wang2009,Takahashi2013}. 
It was suggested that hydrogen atoms chemisorb on Fe$_n$ in the first step.  Additional hydrogen atoms are attached by physisorption in the second step.  Therefore, complexes and supercomplexes containing one, or a few, Fe atoms and  many H atoms are formed.  
We study hydrogenated iron nanoparticles to shed light on the missing cosmic iron problem.

\section{Computational methods}
\label{methods}

We use density functional theory (DFT) methods \citep{Becke2014,Jones2015}. In DFT the total energy of a quantum system is determined by the electron density $n({\vec r})= \Sigma |\psi ({\vec r)}|^2$ of the ground state. By using various physical and computational algorithms DFT gives a realistic description of many physical and chemical properties of nuclei, atoms, molecules, nanoparticles, and solids.

Results are obtained by  
the GPAW code \citep{Enkovaara2010} and its ASE user interface \citep{Bahn2002}. The generalized Gradient Approximation (GGA) \citep{Perdew1997} in its spin-polarized form is chosen, as well as the PAW pseudopotentials \citep{Mortensen2005}.
We start from  Fe-H nanoparticles first optimized  globally,  and then using GPAW DFT methods, by  Takahashi and coworkers \citep{Takahashi2013,Takahashi2014}.
However, we use the newest version of the GPAW package and reoptimize structures for the grid spacing of 0.14 \AA. 
Takahashi and coworkers investigated structural properties and  bonding  of Fe-H nanoparticles. 
They also reported magnetic moments for a few of Fe-H clusters.
We are interested in astrophysical applications. In this work we calculate infrared (IR) spectra, electric and magnetic moments of hydrogenated iron nanoparticles. 
IR spectra are obtained using the finite-difference approximation for a dynamical matrix and the gradient of the dipole momentum 
\citep{Porezag1996,Frederiksen2007}. 
The IR intensity $I_i$ of the mode $i$ is obtained using $$I_i = \frac{N\pi}{3c} \left |\frac{d{\vec{\mu}}}{dQ_i} \right |^2,$$  where
$N$ is the particle density, c is the velocity of light, $\mu$ is the electric dipole momentum, and Q$_i$ is the coordinate of the normal mode.
The same DFT code and similar computational methods in calculations of IR spectra of cement (Ca-Si-O-H) nanoparticles with astrophysical applications\citep{Bilalbegovic2014} produced good agreement with measured IR spectra of the cement paste \citep{Garbev2007}.
The IR frequencies we calculate correspond to stretching and bending modes in
nanoparticles. Calculated data are absorption spectra and they are convolved with the Lorentzian band profiles to adjust them to astronomical emission spectra \citep{Bauschlicher2010}. These vibrational modes are excited when nanoparticles are heated by starlight and their main excitation mechanism is the absorption of a UV  photon. Excited nanoparticles relax by IR emission. However, hydrogenated iron nanoparticles could also form or  adsorb onto dust grains where they could be seen in absorption.

\section{Results and Discussion}
\label{results}

\subsection{IR spectra of hydrogenated iron nanoparticles}

We present here IR spectra for four typical examples of hydrogenated iron nanoparticles shown in 
Figs. 1 and 2.  
FeH$_{10}$ (Fig. 1 (a)) is the 
structure with the largest number of hydrogen atoms connected to the single Fe atom we investigated.
Fe$_9$H$_{56}$ (Fig. 1 (b)) is the hydrogenated iron nanoparticle with the largest number of atoms
we studied. Much larger Fe$_n$H$_m$ nanoparticles, with $n$ up to 130, were investigated in experiments \citep{Parks1985}.
In the Supporting Information
we also show IR spectra for Fe$_3$H$_{25}$  (Fig. 2 (a)) and Fe$_4$H$_{25}$ (Fig. 2 (b)) to compare two structures with the same number of hydrogen atoms bonded to Fe$_n$ and Fe$_{n+1}$.  The Lorentzian profiles with a full width at half-maxiumum  of  20 cm$^{-1}$  are used for 
IR bands presented in Figs. 3 and Fig. S1 in the Supporting Information. 
Tables of bands and their intensities for these structures are available in the Supporting Information.
For some Fe-H nanoparticles the spectrum extends to the far IR. For example, bands of Fe$_4$H$_{25}$  draw out above 100 $\mu$m 
(see Fig. S1 (d) in the Supporting Information).

\cite{Knickelbein1998} carried out an experimental study of multiply hydrogenated iron nanoparticles, F$_n$H$_m$, $n=9-20$. 
IR spectra were recorded in the (9.2-11.3) $\mu$m region. They found that bands overlap and that each band is about 20 cm$^{-1}$   in width. They also carried 
DFT calculations using the local spin density approximation
for only one cluster: Fe$_{13}$H$_{14}$. This structure was constructed to have Th symmetry of the iron core.  Its theoretical IR spectrum was much simpler than measured spectra showing that real hydrogenated iron nanoparticles have low symmetry. In agreement with experimental results of \cite{Knickelbein1998}, we optimize structures with low symmetry and find that their IR bands often overlap.  It  could be difficult to disentangle 
some of these IR features from those of other species in the space. 

\begin{figure}
\centering 
\includegraphics[scale=0.22]{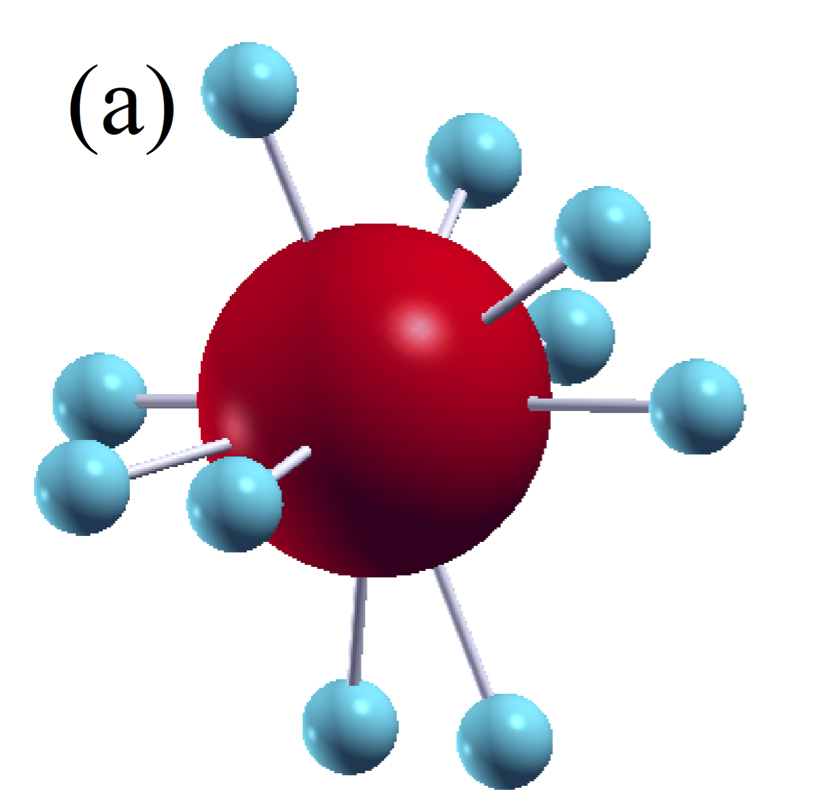}
\includegraphics[scale=0.22]{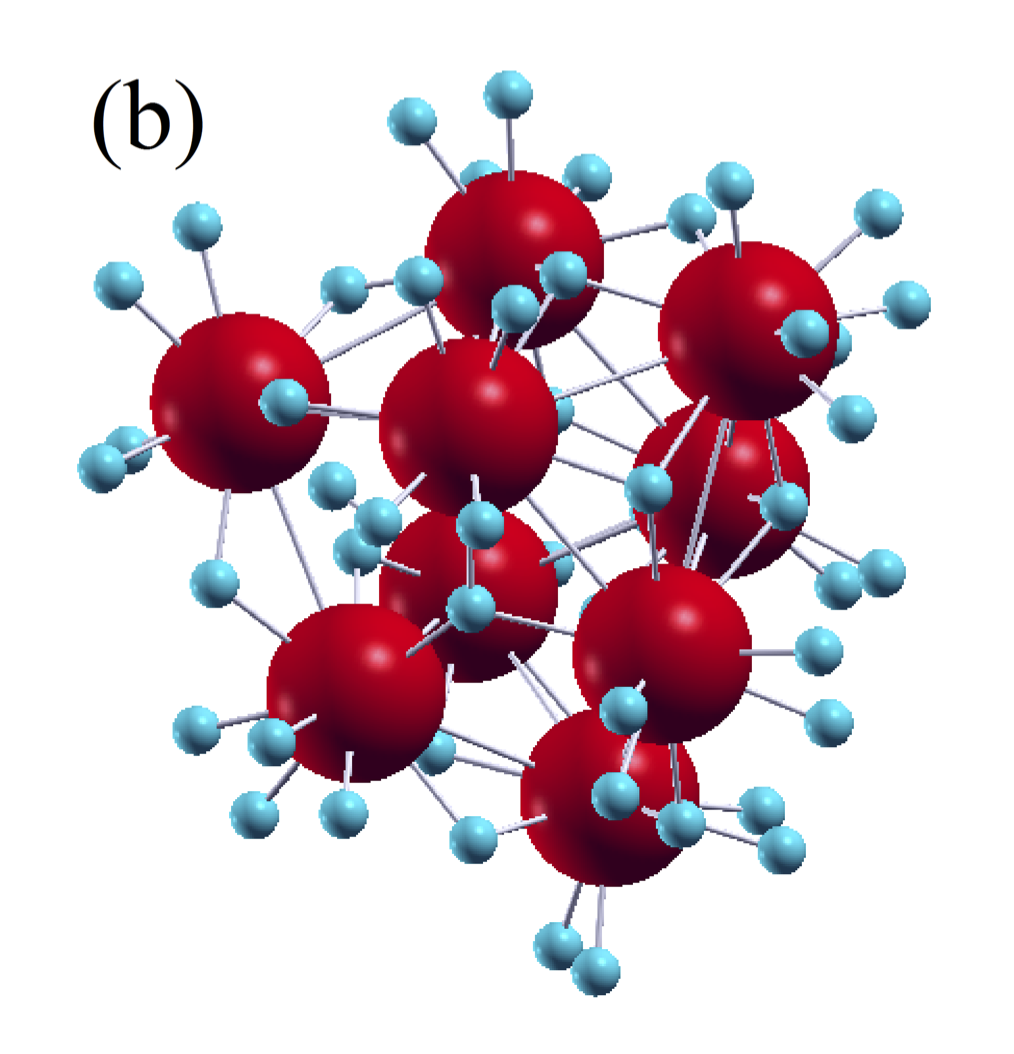}
\caption{ The optimized structures of: 
(a) FeH$_{10}$,
(b) Fe$_9$H$_{56}$. 
Fe and H atoms are represented by the red and blue circles, respectively.}
\label{fig1}
\end{figure}

\begin{figure}
\centering 
\includegraphics[scale=0.22]{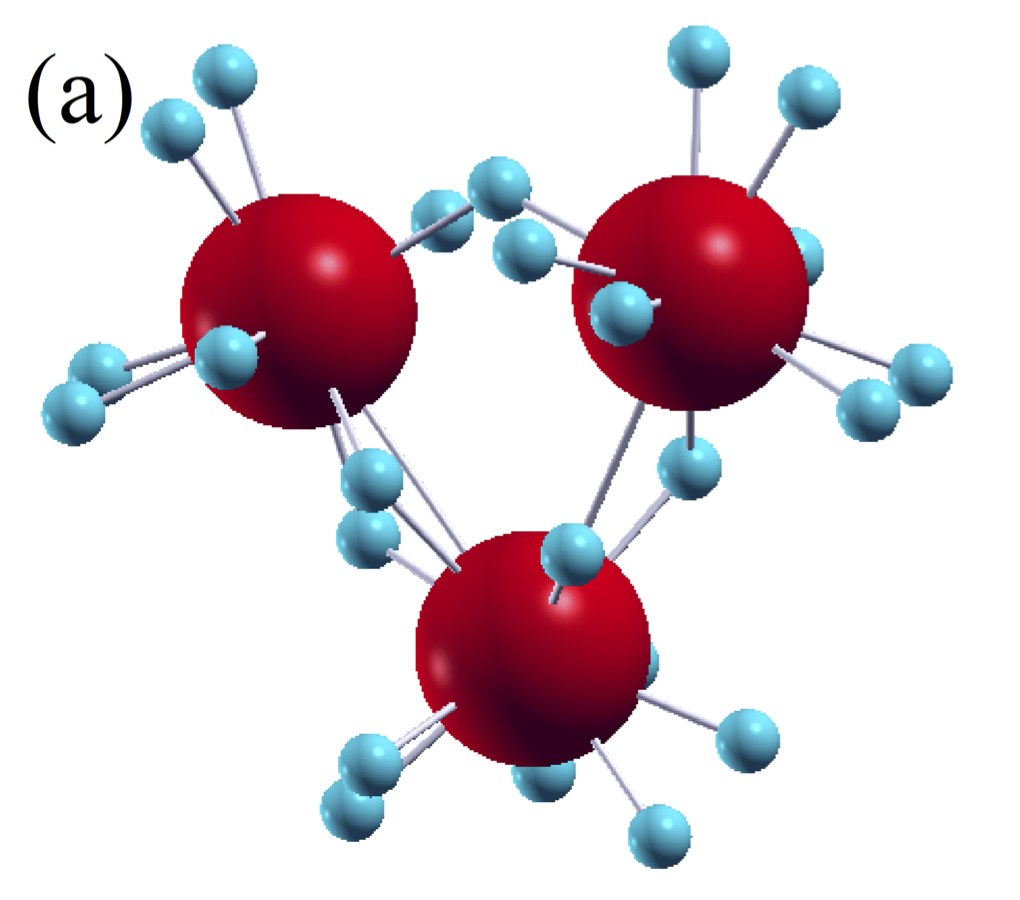}
\includegraphics[scale=0.22]{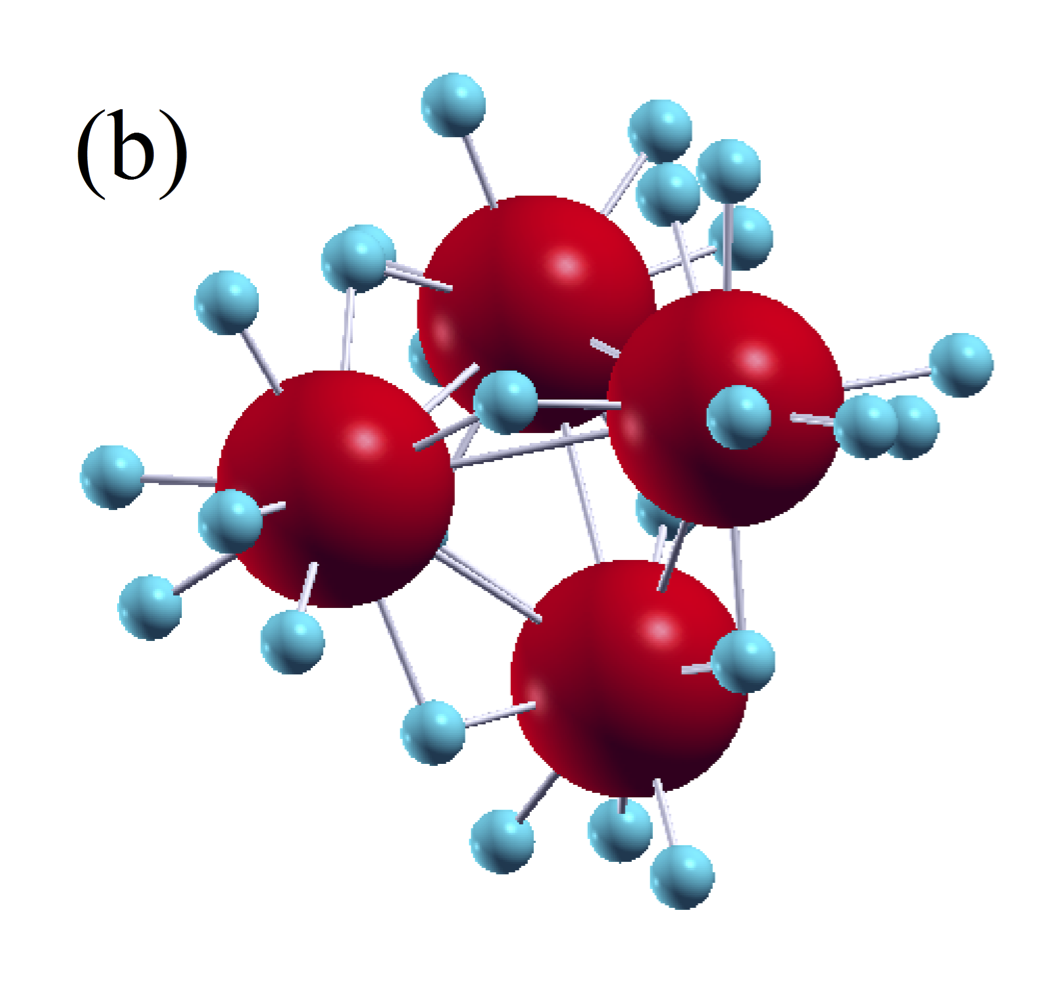}
\caption{ The optimized structures of: 
(a) Fe$_{3}$H$_{25}$,
(b) Fe$_{4}$H$_{25}$. 
 Fe and H atoms are represented by the red and blue circles, respectively.}
\label{fig2}
\end{figure}

\begin{figure}
\centering 
\includegraphics[scale=0.4]{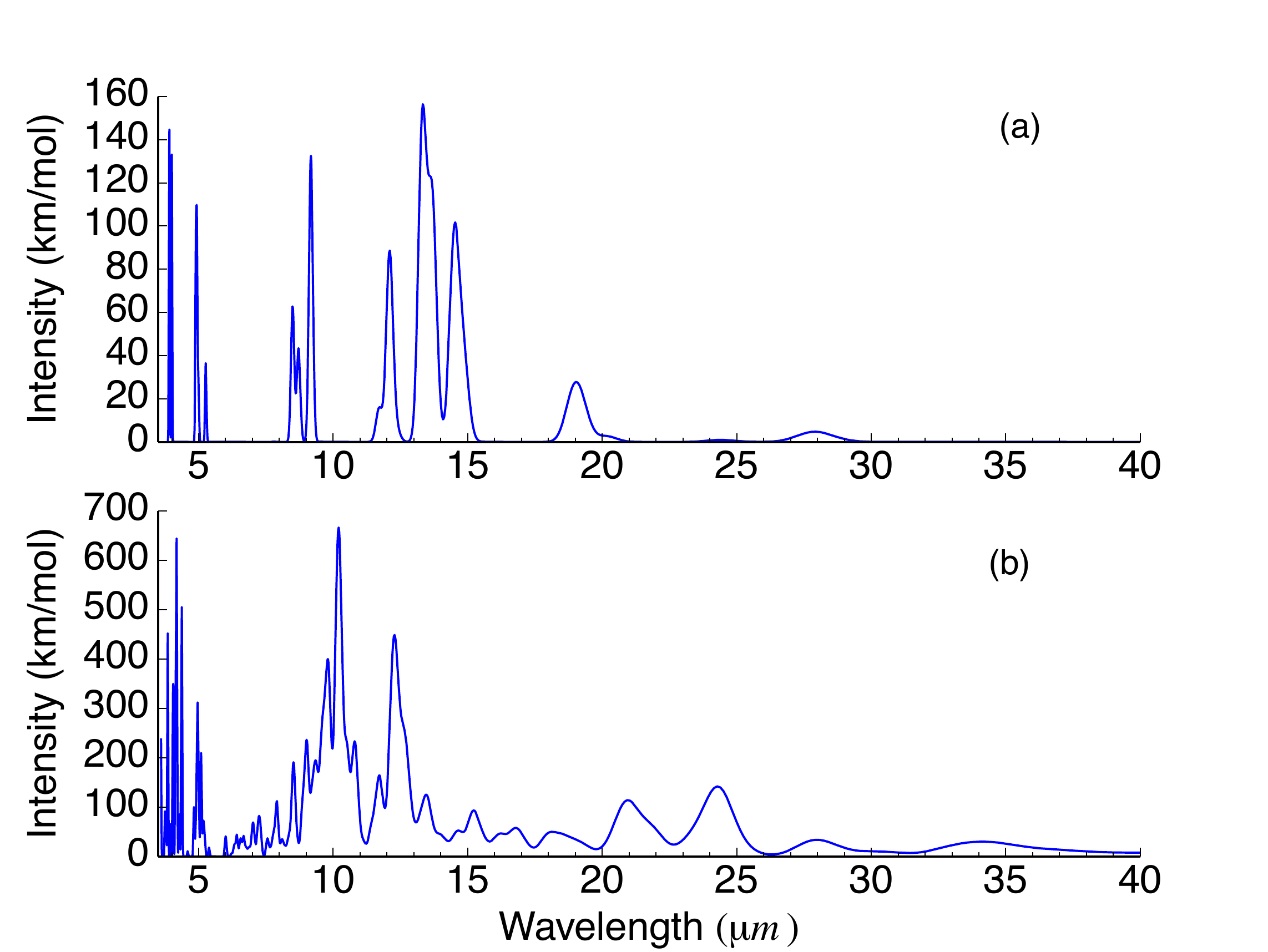}
\caption{ IR spectra of: 
(a) FeH$_{10}$,
(b) Fe$_9$H$_{56}$. }
\label{fig3}
\end{figure}

\subsection{Hydrogenated iron nanoparticles and the interstellar electromagnetic fields}

Some of Fe-H nanoparticles have electric and magnetic moments. For example, we calculate magnetic moments of:  4$\mu_B$ (FeH$_2$),  0 $\mu_B$ (FeH$_{10}$), 0 $\mu_B$ (Fe$_9$H$_{56}$).   Starting from FeH$_2$, magnetic moments for FeH$_m$ decrease with $m$  in the oscillatory way.
For Fe$_8$H$_{51}$ we calculate 1 $\mu_B$. For Fe$_3$H$_{25}$ and Fe$_4$H$_{25}$ (shown in Fig. 2) we find magnetic moments of -1 $\mu_B$and +1 $\mu_B$, respectively. All magnetic and electric dipole moments we calculate are shown in Table 1. The values of electric moments fluctuate between zero and 2.6 D. The values of all magnetic moments in Table 1 are rather low. However, \cite{Knickelbein2002} carried out magnetic measurements on Fe$_n$ and Fe$_n$H$_m$, $n=10-25$, nanoparticles and found that hydrogenation substantially increases magnetic moments of iron clusters for $n \geq 13$. The largest enhancement was measured for $n=13$. All $n$ values between 13 and 18 also show big enhancements. Magnetic moments of  F$_n$H$_m$, $n=13-18$, were (4-5) $\mu_B$, whereas the corresponding values for Fe$_n$ were (2-3) $\mu_B$ \citep{Knickelbein2002}. It is possible to expect that the most pronounced interaction with the magnetic field  occurs for hydrogenated iron dust grains of this size. 
Other intervals  of enhancements for magnetic moments of F$_n$H$_m$  (for $n > 25$) are also possible.

\begin{table}
\centering
\caption{Calculated magnetic (M)  and electric dipole ($\mu$) moments for Fe-H species.}
\label{tab1}
  \begin{tabular}{ l l l}
  \hline
    System                 & M ($\mu_B$) &  $\mu$ (D)  \\ 
  \hline 
  FeH$_2$                 &        4             &     0                 \\   
  FeH$_3$                 &        3            &    2.361                 \\ 
  FeH$_4$                 &        2             &     2.553                \\  
  FeH$_5$                 &        1             &      1.339               \\  
  FeH$_6$                 &        2            &     0.125                \\  
  FeH$_7$                 &        1             &      0.779              \\  
  FeH$_8$                 &        2            &      0.025               \\  
   FeH$_9$                &        1             &      1.040               \\
  FeH$_{10}$            &        0             &     1.232            \\   
  Fe$_2$H$_{18}$    &        0             &      1.510           \\    
  Fe$_3$H$_{25}$    &        1             &     0.675             \\      
  Fe$_4$H$_{25}$    &        1            &     1.012             \\     
  Fe$_5$H$_{28}$    &         0            &     1.218               \\     
  Fe$_6$H$_{32}$    &         0            &     0.611             \\     
   Fe$_7$H$_{39}$   &         1            &   1.429              \\     
   Fe$_8$H$_{51}$   &         1            &      2.324            \\  
   Fe$_9$H$_{56}$   &         0            &     0.454            \\    
   
    \hline
  \end{tabular} 
\end{table}

Therefore, some of Fe$_n$H$_m$  nanoparticles interact with the cosmic electromagnetic field.  The alignment of interstellar grains is well known 
problem \citep{Andersson2015}. Various mechanisms of this alignment have been discussed for some time. Present understanding is that larger dust grains are aligned because of radiative processes, whereas alignment by a paramagnetic relaxation is feasible for small grains. 
The 2175 \AA\, extinction bump was attributed to PAH  (Polycyclic Aromatic Hydrocarbon) molecules \citep{Steglich2011}.
It was found that this 2175 \AA\,  feature shows polarization \citep{Martin1999}. Therefore,  it is important to study magnetic properties and polarization of the smallest dust grains, such as PAHs and Fe-H nanoparticles.  
The alignment of silicate grains with inclusions in the form of iron nanoparticles
was recently investigated \citep{Lazarian2016}. Most of
previous studies on alignment of silicate grains with Fe were based on the model where iron atoms are distributed 
diffusively. It was found that iron nanoparticles inclusions in grains can enhance their alignment. 
The core of iron atoms is also present in the Fe$_n$H$_m$  nanoparticles. 

It is important to consider a possible photodestruction of Fe$_n$H$_m$. 
Similar problems, i.e., loss of aromatic H and side-groups, have been studied for PAHs \citep{Tielens2005}. Reaction rates for these processes have been evaluated within several models (RRK, RRKM, RKM, QRKM). Because of the exponential dependence, small errors in the energies and entropies in unimolecular reactions produce large errors in the reaction rates. Therefore, better solutions are based on experimental data \citep{Tielens2005}. Hydrogenated iron nanoparticles are much less studied than PAHs, for which plenty of data accumulated for a long time in chemistry and physics, as well as
over last thirty years in astrophysics. However, \cite{Parks1987} found  that multiphoton absorption for hydrogenated iron nanoparticles leads primarily to loss of hydrogen adsorbates because of the less energy this process takes in a comparison with breaking a metal-metal bond. One-photon fragmentation was also observed in a few cases.

\cite{Liu1985} found that the laser-induced multiphoton absorption on hydrogenated iron nanoparticles leads to desorption of a specific number of H$_2$ molecules. The desorption energy of 1.3 eV was estimated, but it was also found that this value slightly depends on the coverage.
They also analyzed their experimental results within the RRKM model  \citep{Liu1985}.
In experiments on hydrogen-saturated iron nanoparticles \cite{Knickelbein1998} observed that the laser-induced heating below 150K  leads to a loss of some of the physisorbed H$_2$ molecules. 
UV lasers were used in experiments on photodestruction of Fe$_n$H$_m$ \citep{Parks1987,Liu1985,Knickelbein1998}.
The strength of the  FUV interstellar field is 1.7 \citep{Draine1978}, or 1.6 \citep{Parravano2003},
times greater than the Habing field of 1.6 x 10$^{-3}$ erg cm$^{-2}$ s$^{-1}$. Therefore,
desorption, induced by radiation and heating,  could occur in the ISM. It was suggested that the H$_2$ desorption from small dust grains could induce their spinning \citep{Mathis1986}. 

The anomalous microwave emission (AME)  is  the excess emission observed in the (10-60) GHz frequency range \citep{Kogut1996,Leitch1997,Dickinson2013}.
Electric dipole emission from spinning small dust grains \citep{Draine1998}
and magnetic dipole emission from magnetic nanoparticles \citep{Draine1999}
were proposed as the source of the AME.
PAHs are often studied as such very small grains. 
By analyzing full-sky observations in the IR and microwave region from the Planck Collaboration \citep{Planck2014}, \cite{Hensley2016} found that there is no correlation between fluctuations of the AME intensity and fluctuations in the emission of PAHs. 
Other carriers, such as nanosilicates \citep{Hoang2016}, were recently  proposed for a possible orgin of the AME. Fe-H nanoparticles could be also present in the ISM and their spinning, electric  and magnetic emissions could be one of sources of the AME.
 
\subsection{{\it Astrophysical Origin}}

FeH molecules have been seen in brown dwarfs, S stars and M-type giants till now. Brown dwarfs do not have typical stellar winds and the amount of FeH molecules they could inject to the ISM is small. However, S stars and M giants do have pronounced winds \citep{Ramstedt2009,Ferrarotti2002,Kudritzki1978} and could inject FeH to the ISM. The presence of FeH in S stars and M giants is discussed in \cite{Wing1972,Nordh1977,Clegg1978,Lambert1988,Hirai1992}. 
\cite{Mould1978} calculated column densities of FeH for several models of stellar atmospheres for S and M stars. This calculation was updated by \cite{Lambert1980}. The values at 3000 K are: log N  = 15.3 for the M giant (C/O=0.6), log N = 15.8 for the S star (C/O=0.98),  log N = 15.9 for the S star (C/O=1).

It is known that PAHs are destroyed in the ISM by  shocks  \citep{Micelotta2010b},  
in a hot post-shock gas \citep{Micelotta2010a}, and by cosmic rays \citep{Micelotta2011}.  The PAHs injection time to the ISM is longer than a timescale for their destruction. The same problem exists for interstellar dust grains \citep{Jones1994}. It was suggested that PAHs and grains reform in the ISM.  An efficient process is condensation on already existing grains \citep{Jones1994,Dwek2016}. 
Similar mechanisms play a role in the destruction and  replenishment of hydrogenated iron nanoparticles.

\cite{BarNun1980} suggested that FeH and FeH$_2$ may form on grain surfaces in very cold interstellar clouds. After performing and analyzing laboratory
astrophysical experiments, they proposed that H atoms diffuse  and react with Fe atoms adsorbed on the graphite, silicate  or icy grains at low temperatures. The energy of 1.7 eV released in the chemical reaction Fe + H $\rightarrow$ FeH is absorbed by a grain. The ejection of FeH and FeH$_2$ from grains by cosmic rays and sputtering during cloud collisions was proposed. Hydrogenated iron species could be also ejected from very small grains by the energy released from chemical reactions during the bond formation 
\citep{BarNun1980}. Recent DFT and experimental studies have shown that hydrogenation of Fe nanoparticles is preserved on a single layer graphene supported by the Cu substrate \citep{Takahashi2014b}. Iron nanoparticles were deposited by vacuum deposition and hydrogenation was done with hydrogen gas under 1 atm and at the liquid nitrogen temperature. The copper substrate (on which a graphene layer is physisorbed) was chosen because of commercial applications \citep{Takahashi2014b}. We expect that some other substrates, including cosmic dust grains, show similar properties, i.e. support and preserve hydrogenation of iron nanoparticles under suitable conditions. 
For example, \cite{Navarro2016} 
studied the formation of H$_2$ on Fe-containing olivine-based interstellar grains using DFT methods. They found that H adsorption on Fe sites is much stronger than on Mg ones. Because of the strong chemisorbed Fe-H bond, Fe-containing olivines can capture hydrogen atoms on an astronomical time scale. 
Using computational methods \cite{Fioroni2016}
studied the H$_2$ formation on siliceous surfaces grafted with Fe$^+$.
 It was found that Fe-H and Fe-H$_2$ formation is always thermodynamically favored. 
The H atom remains on the Fe-H center increasing  the probability for a second atom to react. 
Therefore, hydrogenated iron nanoparticles could form at S stars, M giants, and on dust grains in the ISM.

IRC +10216 (CW Leo) is the nearby, asymptotic giant branch evolved carbon star known as one of the brightest mid-IR sources outside the Solar system. 
More than 80 molecular species, as well as many  unidentified lines, 
have been detected in  IRC +10216 \citep{Cernicharo1996,Monnier2000,Mauron2010,Tenenbaum2010,Agundez2012,Cernicharo2013,Gong2015}.
Gas phase iron in significant abundance  was detected  in the circumstellar envelope of IRC +10216 with the UVES spectrograph  \citep{Mauron2010}. Observed column densities of refractory metals indicated that iron and other such elements are not completely removed from the gas phase by dust condensation. The FeCN molecule was detected in IRC +10216 \citep{Zack2011}, as well as several other molecules containing metals. Therefore, we suggest that  IRC +10216 is one of astrophysical environments where hydrogenated iron nanoparticles may form. The upper limit  of 10$^{-9}$ (relative to H$_2$) to the FeH abundance   in IRC +10216  was proposed, under the assumption that the emitting region has an average kinetic temperature of 300 K
\citep{Cernicharo2010}.
\cite{Ozin1984}, as well as \cite{Rubinovitz1986}, found that UV radiation promotes the formation of FeH$_2$. We expect that the formation of  hydrogenated iron nanoparticles is efficient in astrophysical environments with UV radiation.

\section{Conclusions}
\label{concl}

It is very important to study iron and its compounds in order
to understand the properties of cosmic dust and molecules in the ISM.
While pure iron, its oxides and sulfides, were discussed as components of cosmic dust grains, Fe-H nanoparticles were not studied in the astrophysical literature, to the best of our knowledge.
We studied hydrogenated iron nanoparticles to point out their possible role in the balance of iron and hydrogen in the ISM.
Fe atoms are sometimes surrounded with many hydrogen atoms. Therefore,  it is difficult for iron atoms in the core to interact with non-hydrogen atoms.  This could explain why only two molecules containing  iron atoms were detected in the space till now.
We use density functional theory methods and calculate IR spectra of Fe-H nanoparticles, their electric and magnetic moments. 
The IR spectrum of hydrogenated  iron nanoparticles consists of broad, overlapping bands that could be difficult to separate from those of other species in the ISM.  We propose observations in the radio, optical and UV spectral regions as additional, and perhaps 
better, tests for detection of hydrogenated iron nanoparticles in the ISM.
Hydrogenated iron nanoparticles with electric and magnetic moments interact with cosmic electromagnetic fields. In addition, H$_2$ molecules could desorbe from highly hydrogenated nanoparticles  yielding to their spinning. Therefore, hydrogenated nanoparticles are good candidates for the analysis of processes such as the alignment of interstellar grains and the anomalous microwave emission. We suggest S stars and M giants,  as well as
IRC +10216, as astrophysical sources where the search for hydrogenated iron nanoparticles could start.
New  experimental, theoretical,  and observational studies  are needed to reveal a role of hydrogenated iron nanoparticles in the ISM.

\section*{Acknowledgments}
This work  was done using computer resources at the University of Zagreb Computing Centre SRCE and at
the Rudjer Bo\v skovi\' c Institute.
GB acknowledges the support of the HRZZ grant IP-2014-09-8656 ``Stars and dust: structure, composition and interaction'', as well as of the QuantiXLie Center of Excellence. We are grateful to the anonymous referee for useful comments.
This research has made use of NASA's Astrophysics Data System Bibliographic Services.

\bibliographystyle{mn2e} 
\bibliography{irfe}

\newcommand{\beginsupplement}{%
        \setcounter{table}{0}
        \renewcommand{\thetable}{S\arabic{table}}%
        \setcounter{figure}{0}
        \renewcommand{\thefigure}{S\arabic{figure}}%
     }

\beginsupplement

\onecolumn{
\begin{center}
\textbf{{\Huge Supporting Information }}
\end{center}
}

\clearpage

\begin{table*}
\centering
\caption{Infrared bands and intensities for FeH$_{10}$.  Only bands with intensities above 2 km mol$^{-1}$ are shown.}
\label{tabS1}
  \begin{tabular}{ l l}
  \hline
    Band ($\mu$m)  & Intensity (km mol$^{-1}$)  \\ 
  \hline 
  27.907      &     4.740                 \\    
  20.185      &     2.570                \\   
  19.197      &    13.425                \\   
  19.064      &     2.969               \\      
  18.885      &    14.827                \\     
  14.855      &    34.916               \\     
  14.508      &    95.236               \\     
  13.684      &   109.959              \\     
  13.317     &    148.760              \\     
  12.376     &      4.460             \\ 
  12.097     &     88.331              \\ 
  11.709     &     15.564             \\
    9.172    &     132.600             \\
    8.713    &      43.290            \\
    8.493    &      62.783            \\
    5.290    &        4.042           \\ 
    5.263    &        34.319          \\  
    4.994    &      26.525            \\
    4.975    &        3.186           \\ 
    4.936    &       91.542            \\
    4.896    &       84.764           \\  
    4.004    &      133.180          \\  
    3.922    &      100.403          \\  
    3.902    &       92.107          \\  
    \hline
  \end{tabular} 
\end{table*}

\begin{table*}
\centering
\caption{Infrared bands and intensities for Fe$_9$H$_{56}$. Only bands with intensities above 2 km mol$^{-1}$ are shown.}
\label{tabS2}
  \begin{tabular}{ l l l l l l l l}
  \hline
    Band ($\mu$m)   & Intensity   (km mol$^{-1}$) &    Band ($\mu$m)   & Intensity   (km mol$^{-1}$) &     Band ($\mu$m)   & Intensity   (km mol$^{-1}$) &     Band ($\mu$m)   & Intensity   (km mol$^{-1}$)\\   
  \hline 
     70.484     &     2.714       &    15.755    &     2.604            &  10.034       &       53.319             &      6.332        &    23.579          \\                                       
     57.020     &     7.408       &    15.572    &     23.439           &  9.895        &       96.836             &      6.228        &     6.793            \\             
      50.455    &     4.871       &     15.359   &    12.316            &  9.820        &      234.609             &      6.005        &    41.042           \\                        
       46.298   &     4.343       &     15.213   &    61.124            &  9.747        &      110.974             &      5.387        &    18.220            \\                
    45.221      &      4.878      &     15.109   &    19.163            &  9.661        &      115.676             &      5.231        &     2.832            \\                
     45.013     &     12.676     &     14.821    &      5.943           &  9.628        &       45.308             &      5.227        &    39.017            \\                
    42.249      &     3.256       &   14.714     &     20.707           &  9.566        &       26.118             &      5.188        &     4.913         \\                
    40.093      &      4.512      &   14.659     &      9.323           &  9.554        &      113.140             &      5.178        &    60.208         \\                
     37.083     &     5.313       &    14.518    &     22.481           &  9.482        &       23.876             &      5.147        &     7.977         \\                
      36.768    &    4.862        &    14.326    &     3.868            &  9.382        &      116.885             &      5.112        &     6.839         \\                
        34.721  &   20.905       &     14.220    &     5.266            &  9.326        &       19.232             &      5.096        &    94.773        \\                 
     33.297     &    14.692      &     14.054    &    20.588            &  9.291        &       68.690             &      5.090        &   101.302         \\                
      32.892    &     4.518       &     13.983   &   11.080             &  9.203        &       83.438             &      5.079        &    10.908         \\                
    30.462      &   8.696         &     13.836   &    4.819             &  9.109        &       22.454             &      5.065        &     4.064         \\                
    29.129      &     3.976       &      13.792 &    16.252             &  9.022        &      105.420             &      5.003        &   100.868         \\                
     28.368     &    11.293      &       13.569 &    10.434             &  9.009        &      110.487             &      4.975        &   54.379         \\                 
    27.788      &    23.881      &       13.523 &     53.069            &  8.869        &      105.885             &      4.971        &  113.044          \\                
    25.232      &    10.288      &       13.441 &   48.453              &  8.664        &       24.228             &      4.953        &  140.436          \\                
    24.425      &      84.131    &       13.339 &    4.836              &  8.543        &      116.443             &      4.934        &   45.315          \\                
     24.179     &    50.943      &        13.271&   16.988              &  8.503        &       76.369             &      4.906        &   48.823          \\                
    23.539      &     28.586     &       13.182 &    38.362             &  8.455        &       10.287             &      4.900        &   13.022          \\                
    23.103      &   14.36 1      &       12.928 &    22.052             &  8.363        &       26.585             &      4.879        &   11.338          \\               
    22.050      &     10.682     &       12.886 &   21.105              &  8.334        &       11.198             &      4.828        &  100.180         \\                
    21.906      &    31.207      &       12.747 &     59.203            &  8.213        &       16.082             &      4.593        &    11.067        \\                
   21.473       &    25.968      &       12.723 &     29.621            &  8.095        &       32.949             &      4.377        &  345.287         \\                
    20.984      &     79.193     &       12.655 &     89.682            &  7.919        &       32.797             &      4.372        &   167.467        \\                
   20.616       &    25.595      &        12.580&     49.768            &  7.907        &       77.750             &      4.295        &   53.979         \\                
   20.455       &   16.582       &        12.487&    28.215             &  7.785        &       42.381             &      4.288        &   10.133         \\                
   19.903       &    3.373        &      12.419 &     100.261           &  7.681        &       13.564             &      4.273        &   44.411          \\               
   19.520       &    5.584        &      12.317 &     137.383           &  7.557        &       36.381             &      4.194        &  222.370           \\               
   19.255       &   7.765         &       12.240 &     192.146          &  7.296        &       50.272             &      4.180        &   392.563        \\                
   19.088       &    14.444      &        12.161 &      67.349          &  7.224        &       60.498             &      4.176        &   88.638           \\               
   18.678       &     13.223     &        12.095 &      62.937          &  7.189        &        3.032             &      4.147        &  278.499          \\               
   18.564       &     16.996     &        11.784  &     93.512          &  7.143        &        6.248             &      4.133        &   76.476           \\              
   18.271       &   10.651       &        11.661  &       92.412        &  7.085        &       15.887             &      4.050        &  352.226           \\              
    18.133      &  12.718        &        11.433  &      58.935         &  7.024        &       54.520             &      3.946        &   65.796           \\              
  17.936        &    6.649        &      11.113   &      27.914         &  6.973        &       18.399             &      3.866        &    98.050            \\             
 17.890         &    22.082      &       10.854   &      150.226        &  6.885        &       14.362             &      3.854        &  169.623             \\            
  17.238        &    7.295        &      10.737   &     124.702         &  6.844        &        5.448             &      3.849        &    238.873           \\             
  17.088        &   11.490       &       10.511   &      204.345        &  6.768        &       16.451             &      3.793        &     69.196          \\              
  16.858        &   6.842         &      10.292   &     337.588         &  6.673        &       40.457             &      3.751        &     91.593          \\              
 16.765         &   40.762       &        10.181  &     327.461         &  6.574        &       29.044             &      3.601        &    237.929          \\              
  16.333        &    20.843      &        10.159  &      59.929         &  6.529        &       13.808             &                   &            \\              
16.099          &     28.915     &        10.108   &    146.893         &  6.422        &       43.568             &                   &           \\

    \hline
  \end{tabular} 
\end{table*}

\begin{table*}
\centering
\caption{Infrared bands  and intensities for Fe$_3$H$_{25}$. Only bands with intensities above 2 km mol$^{-1}$ are shown.}
\label{tabS3}
  \begin{tabular}{ l l l l}
  \hline
    Band ($\mu$m)   & Intensity  (km mol$^{-1}$)  &  Band ($\mu$m)   & Intensity  (km mol$^{-1}$)\\ 
  \hline 
   83.662   &     2.516   &  10.836         &  53.799    \\  
   56.303   &     9.143   &    9.986        &   15.568    \\   
   49.230   &    18.609   &    9.684        &  197.290   \\  
   43.906   &    18.599   &    9.614        &   26.290   \\  
   39.161   &    11.289   &    9.495        &  122.583   \\  
   35.955   &     4.639   &    9.153         &  48.275   \\  
   35.075   &     6.355   &    9.116        &  153.497   \\   
   32.489   &    24.243   &    8.896        &   24.220   \\   
   31.354   &    16.930   &    8.827        &   40.978   \\  
   30.722  &      3.996   &    8.729        &  160.651   \\  
   29.519  &     33.188   &    8.699         &  59.320   \\  
   28.926  &      2.998   &    8.243         &  50.655   \\  
   27.174  &      3.229   &    7.636         & 130.980   \\  
   25.855  &     17.867   &    7.179         & 15.297    \\  
   24.742  &      9.958   &    7.032         & 118.927   \\  
   24.311  &      4.998   &    6.935         & 231.976   \\  
   23.741  &      5.680   &    6.688        &  237.549   \\  
   22.184  &     10.948   &    6.188        &    5.885   \\  
   21.612  &     14.462   &    5.854        &    4.895   \\  
   18.847   &    20.441   &    5.809         &  23.723   \\  
   18.111   &    71.394   &    5.639        &   10.266   \\  
   17.016   &    55.199   &    5.536        &   61.879   \\  
   16.912  &     25.146   &    5.496       &    11.737   \\  
   16.472  &     31.744   &    5.424       &     7.760   \\  
   16.088  &     19.281   &    5.378        &    3.222   \\  
   15.619  &     32.282   &    5.360        &    3.794   \\  
   15.427   &    56.003   &    5.214       &     2.513   \\  
   15.080   &    46.897   &    5.100        &   67.315   \\  
   14.836   &    11.827   &    5.053       &    40.754   \\   
   14.388   &    55.554   &    4.061       &   144.049   \\  
   14.370   &    96.957   &    4.048       &   186.391   \\  
   14.133   &    56.047   &    4.023        &  100.385   \\  
   13.523   &     4.067   &     3.765      &    194.473  \\  
   13.415   &    82.563   &     3.653      &    124.806  \\  
   12.871   &    10.453   &     3.645        &  144.024   \\  
   12.588   &    92.607   &    3.635        &   86.744    \\   
   12.106   &    13.944   &     3.498       &   120.231   \\  
   11.935   &    17.547   &     3.479       &   145.826  \\  
  10.991    &    23.095   &                 &    \\

    \hline
  \end{tabular} 
\end{table*}

\begin{table*}
\centering
\caption{Infrared bands  and  intensities for Fe$_4$H$_{25}$. Only bands with intensities above 2 km mol$^{-1}$ are shown.}
\label{tabS4}
  \begin{tabular}{ l l l l}
  \hline
    Band ($\mu$m)   & Intensity  (km mol$^{-1}$) &    Band ($\mu$m)   & Intensity  (km mol$^{-1}$) \\ 
  \hline 
   107.133     &      44.221    &    9.791   &   134.812       \\    
    68.859     &       8.595    &    9.682   &    10.633       \\    
     44.475    &      15.409    &   9.255    &  180.494        \\    
     43.580    &       8.040    &   9.192    &   73.073       \\    
     41.612    &      16.569    &   9.078    &   18.873       \\    
     36.492    &       6.366    &   8.962    &   24.004       \\    
     34.547    &       8.378    &   8.879    &   32.242       \\     
     33.409    &       9.230    &   8.773    &   46.272       \\    
     30.858    &      24.975    &   8.711    &   46.673       \\   
     29.952    &       8.000    &   8.489    &   17.769       \\   
     29.406    &       4.173    &   8.339    &     4.395       \\   
     27.359    &       2.838    &   8.047    &   22.475        \\  
     25.824    &      23.120    &   7.172    &    7.806        \\  
     24.781    &      45.523    &   7.136    &   15.999        \\  
     22.243    &     26.893     &   7.062    &   48.070        \\   
     21.832    &      8.105     &   6.671    &   21.170       \\   
     21.234    &       8.517    &   6.526    &    3.218       \\   
     20.447    &      24.236    &  6.362     &    4.361   \\   
     19.856    &      32.659    &   5.740    &    2.892    \\  
     19.539    &       5.709    &   5.721    &   21.132    \\  
     18.800    &      33.036    &   5.564    &    7.898     \\  
     18.182    &      23.299    &   5.383    &    10.790        \\   
     17.099    &       6.212    &   5.360    &      7.641       \\   
     16.064    &       96.373   &   5.356    &     26.395       \\   
     15.684    &       24.778   &   5.229   &      11.211      \\   
     15.485    &       35.924   &   5.191     &    36.792       \\   
     14.854    &       2.467    &   5.132     &   144.669       \\   
     14.744    &      41.193    &   5.116     &     7.849        \\  
     14.180    &      87.340    &   5.080     &    30.734        \\  
     13.748    &      16.051    &  4.990      &   140.327        \\  
     13.676    &      18.555    &   4.045     &     88.070      \\   
     13.444    &       2.808    &   3.933     &    159.488      \\   
     13.185    &      42.963    &   3.894     &    125.524      \\   
     12.679    &     100.206    &   3.870      &   171.262       \\   
     12.464    &        10.310  &   3.753      &   127.317     \\   
     10.813    &        30.523  &   3.721      &   158.850     \\   
     10.407    &       254.551  &   3.673      &   352.667     \\      
    \hline
  \end{tabular} 
\end{table*}

\clearpage

\begin{figure*}
\centering 
\includegraphics[scale=0.6]{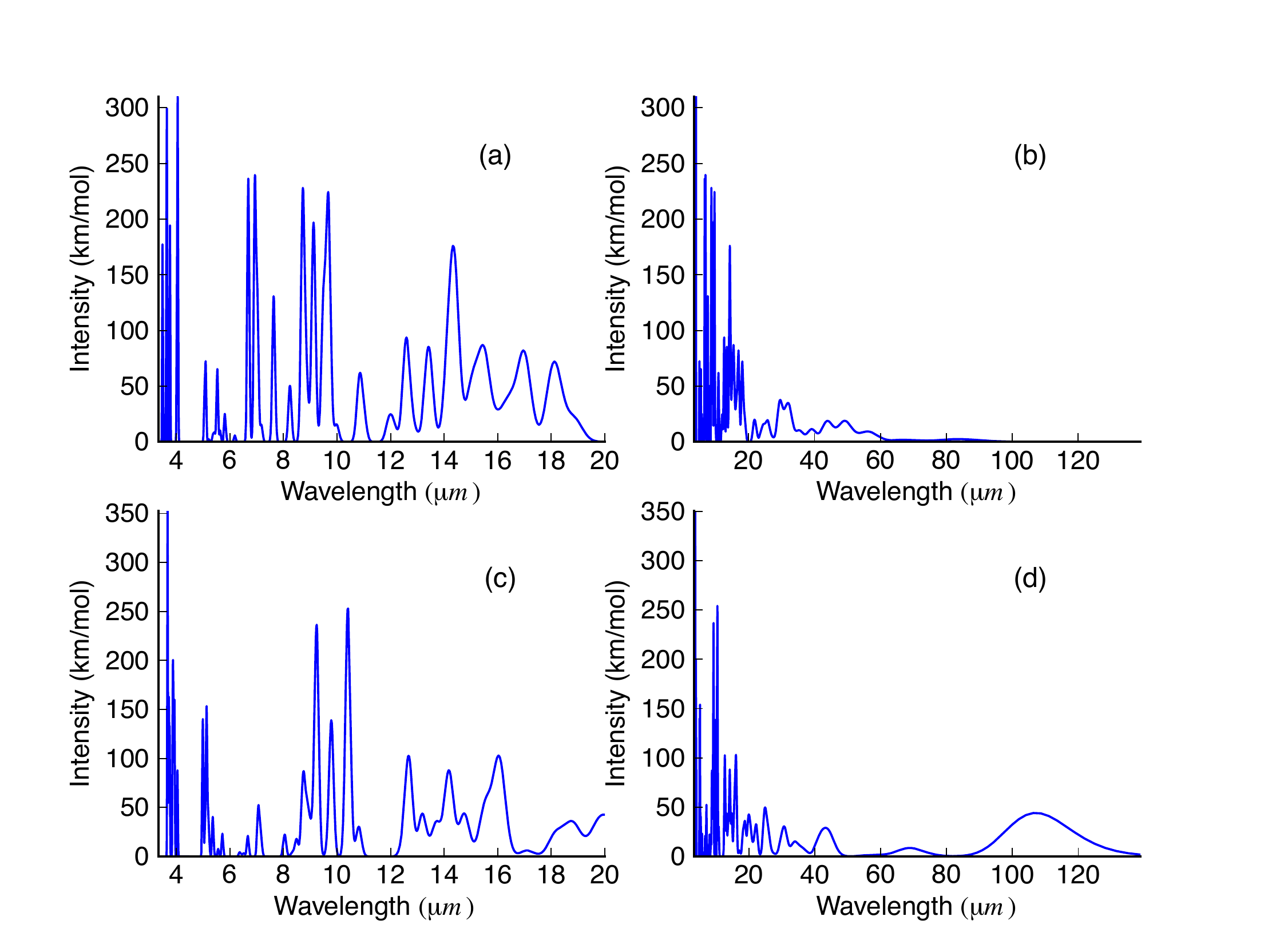}
\caption{ IR spectra of: 
(a) Fe$_3$H$_{25}$,
(b) Fe$_3$H$_{25}$ including far-IR,
(c) Fe$_4$H$_{25}$,
(d) Fe$_4$H$_{25}$ including far-IR. }
\label{fig4}
\end{figure*}

\clearpage

\end{document}